\documentstyle[12pt]{article}
\input psfig
\newcommand{\ga}{\gamma}
\newcommand{\th}{\vartheta}

\newcommand{\De}{\Delta}
\newcommand{\la}{\lambda}

\newcommand{\eps}{\varepsilon}
\newcommand{\om}{\omega}
\newcommand{\k}{\kappa}
\newcommand{\barr}{\begin{array}}
\newcommand{\bea}{\begin{eqnarray}}
\newcommand{\beq}{\begin{equation}}
\newcommand{\ear}{\end{array}}
\newcommand{\eea}{\end{eqnarray}}
\newcommand{\eeq}{\end{equation}}

\newcommand{\bma}{\begin{displaymath}}
\newcommand{\ema}{\end{displaymath}}
\newcommand{\qp} {quasiparticle}

\begin{document}
\begin{center}
{\bf RESTORATION OF THE BROKEN D2-SYMMETRY 
IN THE MEAN FIELD DESCRIPTION OF ROTATING NUCLEI }
\end{center}

\sf

\vspace*{0.2cm}

\begin{center}
F.~D\"onau, Jing-ye Zhang,$^*$ and L.L. Riedinger$^*$
\end{center}
\begin{center}

Institut f\"ur Kern- und Hadronenphysik, FZ Rossendorf, 
01314 Dresden\\
$^*$ Department of Physics and Astronomy,  
University of Tennessee, TN 37996
\end{center}

\date{\today}

\begin{abstract}
{\sf  Signature effects observed in rotational bands 
are a consequence of an inherent D2-symmetry. 
This symmetry is naturally broken by the mean field cranking 
approximation when a tilted (non-principal) axis orientation
of the nuclear spin becomes stable. The possible tunneling 
forth and back between the two 
symmetry-related minima in the double-humped potential-energy 
surface appears as a typical bifurcation of the rotational band. 
We describe this many-body process in which 
all nucleons participate  
by diagonalizing the nuclear Hamiltonian within a selected set of 
tilted and non-tilted cranking quasiparticle states.  
This microscopic approach is able to restore the broken 
D2 symmetry and reproduce the 
quantum fluctuations between symmetry-related
HFB states which emerge as splitting of the band energies and
in parallel staggering in  intraband M1 transitions.}
\end{abstract}

\renewcommand{\baselinestretch}{1.3}\large\normalsize
{\sf 

{\bf 1. Introduction}

The Tilted Cranking (TAC) Model \cite{tac} has been  
a rather fruitful mean field approach
to treat the angle degree of freedom connected 
with non-principal axis orientations of the rotational 
axis\footnote{This axis is defined by the direction of 
the expectation value $\langle$\mbox{\boldmath$I$}$\rangle$ 
of the nuclear spin vector \mbox{\boldmath$I$} for a given state.}  
in deformed nuclei. 
This mean-field theory resulted in the microscopic 
description of $\De I$=1 rotational
band structures, in particular, those which manifest themselves 
by strong magnetic dipole transitions observed in several mass regions.
However, the tilted-axis
rotation implies unavoidably the spontaneous breaking of the   
signature symmetry, i.e.\ the familiar D2 symmetry 
in deformed nuclei with respect to a 180 degree rotation
is lost. 
Here we consider the consequences 
of this symmetry breaking and a possible  
way of its restoration for a typical example,   
the  $K=7/2$ ground-state band in $^{175}$Hf. 
The situation is illustrated in fig.\ 1 which shows the dependence 
of the potential energy surface (PES) on the tilt angle $\th$
for three frequencies as obtained from a TAC calculation.
Due to the D2 symmetry of the deformed density, these PES 
are mirror symmetric. The potential minima   
determine the stable rotational axes of uniformly rotating 
selfconsistent quasiparticle (qp) states,
which turn out to be  tilted for the lowest 
and non-tilted for the excited configuration.

\renewcommand{\baselinestretch}{1.0}\large\normalsize

\begin{figure}[h]
\mbox{\psfig{file=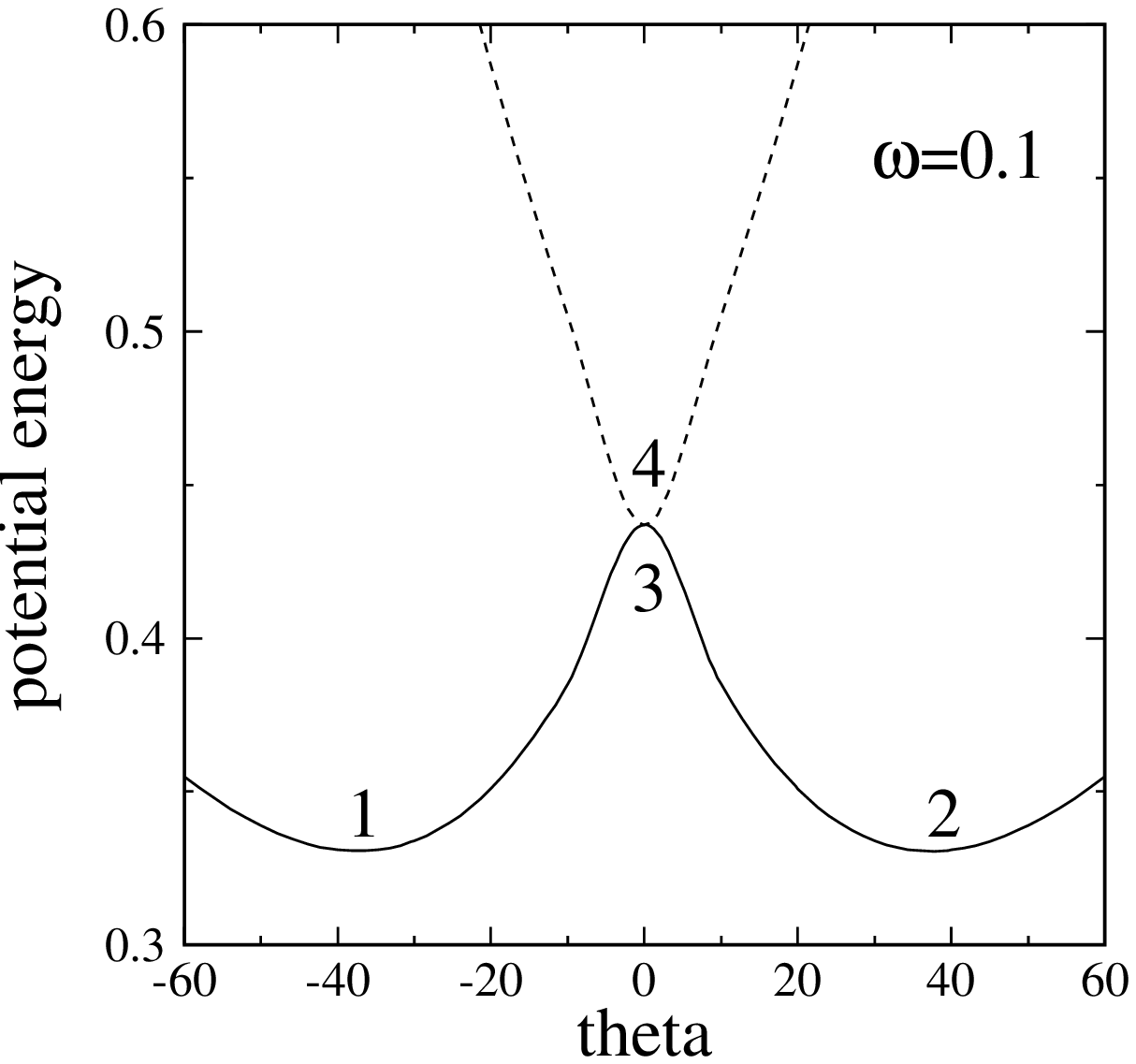,width=4.8cm,angle=0}}
\mbox{\psfig{file=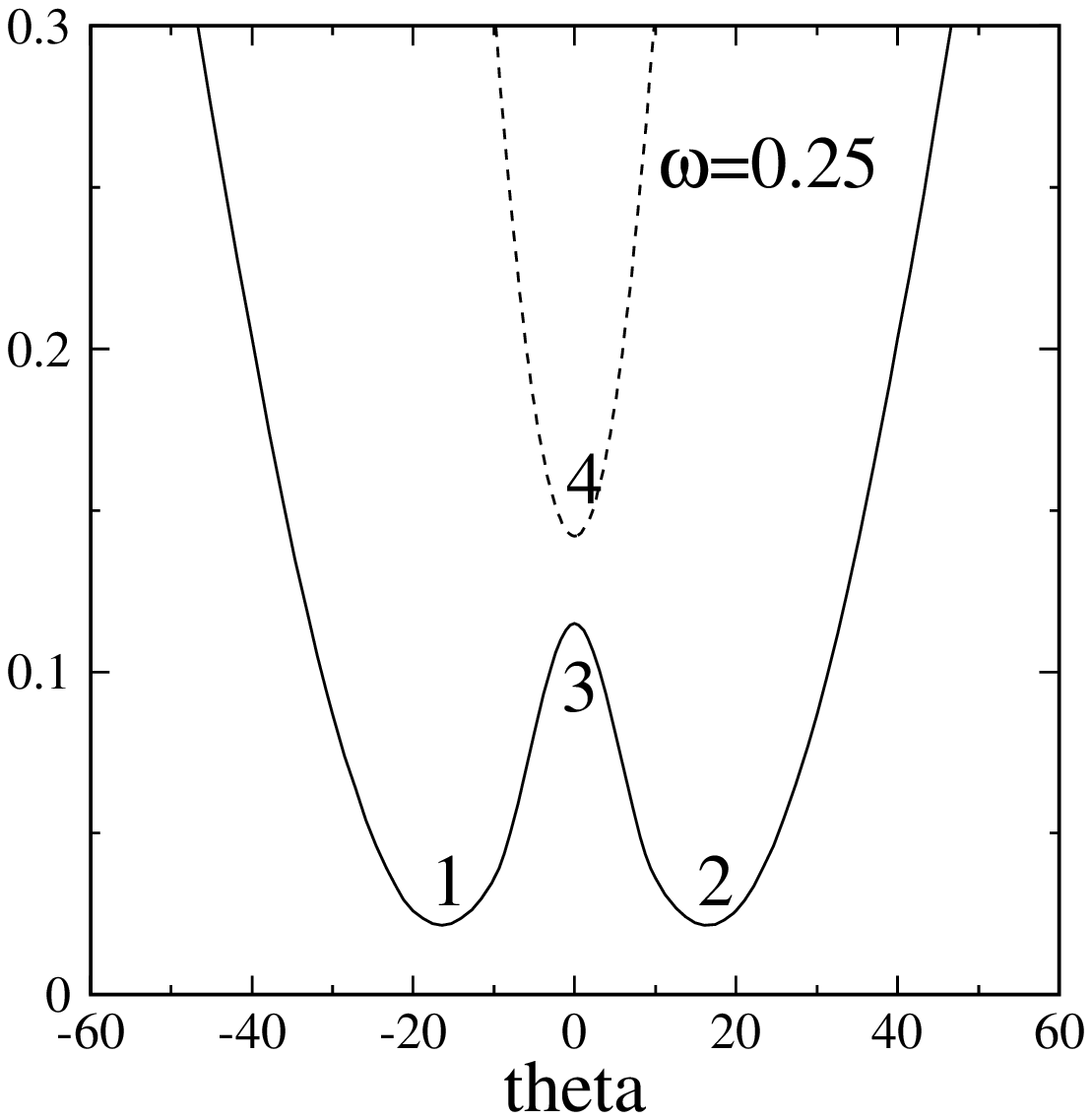,width=4.8cm,angle=0}}
\mbox{\psfig{file=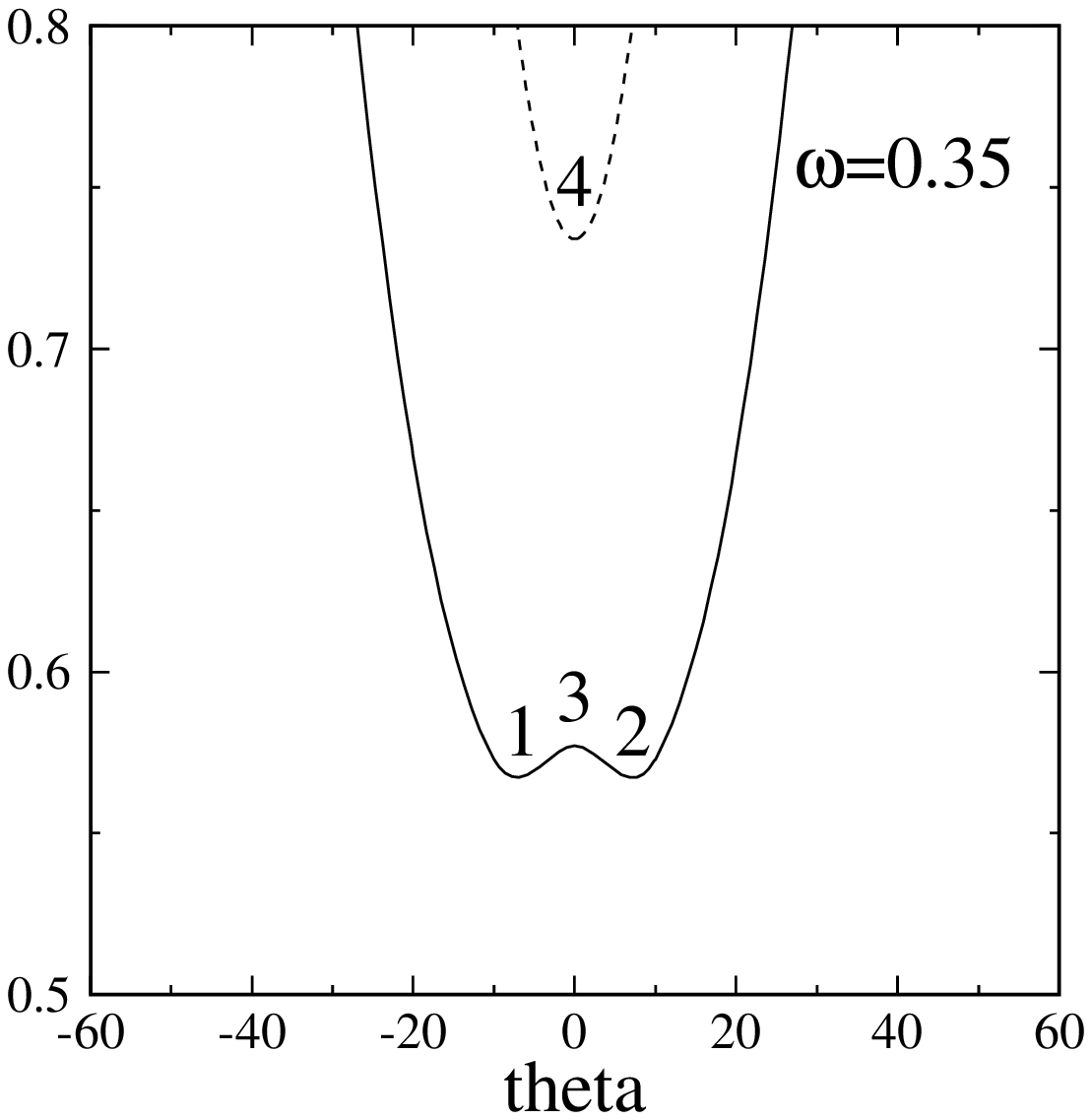,width=4.8cm,angle=0}}
\caption{ \small\it Calculated potential energy surfaces (PES)
at rotational frequencies  $\hbar\om=0.1, 0.25$ and 0.35 MeV
as a function of the tilt angle $\th$ of the cranking axis
with respect to the short (collective) deformation axis.
Solid and dashed lines denote the PES of the lowest and first excited
configuration, respectively.
Note the typical double humped character of the lowest sheet due to
the inherent D2 symmetry of the axially deformed nuclear shape.
The numbers 1-4 correspond to the four extreme points in these PES.
}
\end{figure}

\renewcommand{\baselinestretch}{1.3}\large\normalsize

{\sf

At low frequency there appears two symmetry-related 
minimum points $\th_\circ$ and  $-\th_\circ$
separated by a substantial barrier. 
Correspondingly, there exist two different quasiparticle states
with stably tilted, symmetry-broken spin orientations. In the TAC model
one of these tilted states is selected to represent 
the intrinsic state of the  considered $\Delta I$=1 band. 
This is a good approximation
as long as the possible quantal motion between the different minima  
can be  neglected.
As seen in fig.\ 1  the two equivalent minima approach each other 
with increasing 
rotational frequency and 
 finally merge into a single minimum at $\th$=0, then corresponding 
to the stable rotation about the short, so-called  collective deformation axis. 
The latter is a principal axis 
rotation (PAC) having naturally good signature symmetry.
The onset of the quantal motion
is observed experimentally  as the beginning of the well-known 
$\De I$=1 staggering effect (signature splitting) in the band energies 
 and a corresponding staggering of the B(M1) strength \cite{Hamamoto}
of subsequent $\Delta I$=1 transitions.  

The microscopic description of the signature bifurcation 
goes beyond the concept of the mean field. When  
the spin orientation is artificially fixed along a principal 
deformation axis as done in the
1-dimensional (PAC) cranking model, the signature effect appears as 
 an intrinsic excitation (in fig.\ 1: 3$\rightarrow $4) 
within the same mean field. 
However, this cranking axis happens to be unstable for the lowest 
configuration as seen in fig.\ 1. 
The two-dimensional cranking (TAC)  brings
the spin orientation into a stable but tilted direction which is
manifested nicely in the geometrical spin dependence of the 
$\gamma$ radiation amplitudes \cite{tac}. Hence,
in order to build a bridge between the PAC and the TAC picture, one 
must give up  
the simple determinantal form of the qp 
solutions in favor of an appropriate 
superposition of those states implying both tilted and non-tilted ones. 
It will be demonstrated that a direct diagonalization of the nuclear 
Hamiltonian within a small set of relevant quasiparticle states
provides the desired framework for obtaining the mixing amplitudes
of this superposition.        
This approach, denoted later on as diagonalized superposition 
of quasiparticles
(DISQ), has some features of the Generator coordinate method when identifying
the orientation angle $\th$ as the collective variable in the spirit of 
reference \cite{onishi}.
However, we purposely include also an excited configuration not obtainable
by a continuous change of the angle variable. 
The inclusion of the excited signature partner qp 
state is necessary in order to complete the 
configuration space. At higher frequency in the lowest PES 
only the symmetrical combination of the tilted qp states survives
while merging into the favored signature PAC state, 
whereas the antisymmetric component simultaneously disappears.    
One may relate the signature 
effect to the picture   of quantum tunneling \cite{onishi},
but the tunneling under the barrier seems to play only a partial role. 
The explicit presence of an excited signature-partner state
above the barrier is important as well. 

The practical performance of the DISQ approach is technically ambitious
because the PAC and TAC quasiparticle states are complicated many-body 
states forming in addition a non-orthogonal basis set. The setting up
of the Hamiltonian matrix in such a basis and the subsequent calculation of 
the transition matrix elements were enabled by     
applying here the tools recently developed \cite{canonic} in order
to derive the overlaps and Hamiltonian kernel for non-orthogonal
Hartree-Fock-Bogoliubov (HFB) states. \\

{\bf 2. Hamiltonian}  

The study of the symmetry restoration relies on a 
rotational invariant many-body Hamiltonian which
consists of three parts: a spherical 
average field part with the chemical potential term, 
a residual monopole pairing plus quadrupole interaction,  
and the usual cranking term: 

\bea
H' = h_{sph} -\la N + \sum_{\tau=p,n} G_\tau  P^+_\tau P_\tau
 - \frac{\k}{2} \sum{Q^*_\mu Q_\mu} - \om I_x . 
\eea

The spherical mean field part $h_{sph}$ is taken to be the modified 
oscillator Hamiltonian at zero deformation. The parameters 
$G_\tau$ and $\k$  determine the   
coupling strength of the factorized pairing and quadrupole force, respectively,
where for the latter we assumed the same strength for protons and neutrons.
The last term implying the cranking frequency $\om$ aligns the 
angular momentum
direction (i.e.\ the spin orientation) along the x-axis. 
The self-consistent (s.c.) mean-field solutions corresponding to the above
Hamiltonian $H'$ (because of the cranking term, a Routhian)  are constructed
by means of the following {\it tilted } mean-field cranking Hamiltonian 
denoted also as the qp Routhian:
\bea
&h' = h_{sph} - \lambda N + \sum_{\tau=p,n} \De_\tau (P^+_\tau + P_\tau)&
\nonumber\\
&   - \hbar \om_o  \beta (\cos\ga Q_0 + \sin\ga /\sqrt{2} (Q_2+Q_{-2}))
 - \om ( \cos\th I_x + \sin\th I_z ).&
\eea
Here a two-dimensional cranking term,  $- \om ( \cos\th I_x + \sin\th I_z )$,
is needed to obtain intrinsic qp states with a {\it tilted} spin orientation
that is rotated subsequently to the space-fixed x axis of the lab system.   
The pairing gaps $\De_\tau$ and quadrupole deformation parameters 
$(\beta,\ga) $ are determined by the usual selfconsistent conditions
\bea
& \De_\tau = G_\tau < P_\tau >& \nonumber \\
&\hbar\om_o \beta \cos\ga = \k < Q_0 >, \quad
\hbar \om_o  \beta \sin\ga /\sqrt{2} = \k < Q_2 >,&\\
& < Q_1> = < Q_{-1}> = 0 .&\nonumber
\eea

where the $< Q_{\mu =0,\pm 1,\pm 2} >$ and $< P_{\tau =\pm 1/2} >$ 
are the expectation values with respect to the considered qp state.
The eigenstates of the Routhian $h'$ are the Hartree-Fock-Bogoliubov
determinantal states determined by the actual occupation of the
many quasiparticle orbitals involved.

The Hamiltonian $h'$ coincides with the TAC model \cite{tac}
and we construct our qp states with the oscillator 
basis including the proton
shells $N=3,4,5$ and neutron 
shells $N=4,5,6$.
The selfconsistency of the pairing plus quadupole Hamiltonian is 
easily obtained by choosing appropriate values of the  
deformation parameters and pairing gaps and subsequently matching 
the strength constants from  eqs.\ 1 - 3.  
Hence, the tilted quasiparticle states are formed by standard methods.
From a solution of the HFB equation (cf.\ \cite{Ring}), i.e.\ 
$ [h',a^+_i]=e'_i a^+_i $, the quasiparticle operators 
$a^+_i$ and the Routhian energies $e'_i$ are calculated for given set of  
parameters $\beta,\ga,\De,\om,\th $.

The relevant s.c.\ tilt angle $\th_\circ $ is derived from the 
requirement \cite{tac} that the (here two-dimensional) spin orientation
determined by the expectation values $ (<I_x>,<I_z>)$ of the spin operator
\mbox{\boldmath$I$} 
becomes parallel to the cranking direction,
i.e.\ in our two-dimensional case 
\beq
\frac{\om_x}{\om_z} = ctg(\th_\circ) = \frac{ <I_x>}  {<I_z>}.
\eeq
This parallel condition is equivalent to finding the local minimum  
in the  PES (cf.\ fig.\ 1) of the selected qp configuration.  \\

{\bf 3. Selection of quasiparticle states in $^{175}$Hf}

As mentioned above, a similar description of signature-splitting effects 
within a more conventional GCM approach was tried previously 
in ref.\ \cite{onishi}. 
These authors considered a  set of up to 30 \qp\ states 
forming a path of tilt angles through the PES including 
the two minimal points, but all these qp states develop continuously in
the lowest sheet of the PES, i.e.\ the qp states follow
adiabatically  the minimum configuration. 
However, the outcome of this attempt was not really satisfactory. 
The inclusion of an excited 
configuration seems to be crucial for treating the signature effects.
The previous results indicate the situation that more experience 
is needed  
in order to learn more about both the physics of the large amplitude 
collective
motion and the generator coordinate method. Our calculations below
should be also considered as exploratory studies in this 
direction.      

To simplify the task, we consider all mean field parameters in $h'$,
 except the tilt angle $\th$, to be fixed. 
The adopted deformation parameters for the $K=7/2$ band under study 
are $\eps_2=0.258,\,\, \eps_4= 0.$ and $\ga=0 $. The pairing gaps are 
$\De_p=0.75$ MeV  and $\De_n=0.69$  MeV, respectively. 
Aiming in this paper to extend the TAC mean-field approach in the simplest
manner, we include only 
the four tilted s.c.\ qp states  marked in PES of fig.\ 1 
 in the diagonalization of $H'$.
Two of the selected  points correspond to the symmetrical TAC minima 
of the two stably tilted qp states, denoted below as $|\pm\th_\circ\rangle$. 
The third point belongs to the metastable maximum point of the 
favored signature PAC configuration at $\th=0$, and the 
fourth point is placed also at $\th=0$ into the minimum of 
the unfavored signature PAC configuration (fig.~1). 
These four states are
suggested to play the key role for describing the signature-splitting 
effects. Needless to say,  the normal PAC model
is able to describe the signature splitting of non-tilted 
rotational bands quite successfully \cite{signat}. Thus,
it is suggestive to include the above four states which merge
naturally into the usual PAC signature partners  at higher frequency.
A systematic study for a larger set of points will be done
elsewhere. 

By construction, the qp states of the Routhian $h'$ are intrinsic states
which belong to different spin orientations $\pm\th_\circ$. Therefore,
before the final 4$\times$4 diagonalization of $H'$ 
is done, one has to transform all intrinsic qp states 
to a $\it common $  lab system  via an appropriate rotation.
In this respect, we remind the reader that the spin orientation of a rotational
invariant Hamiltonian is a space-fixed vector.

\begin{figure}[h]
\mbox{\psfig{file=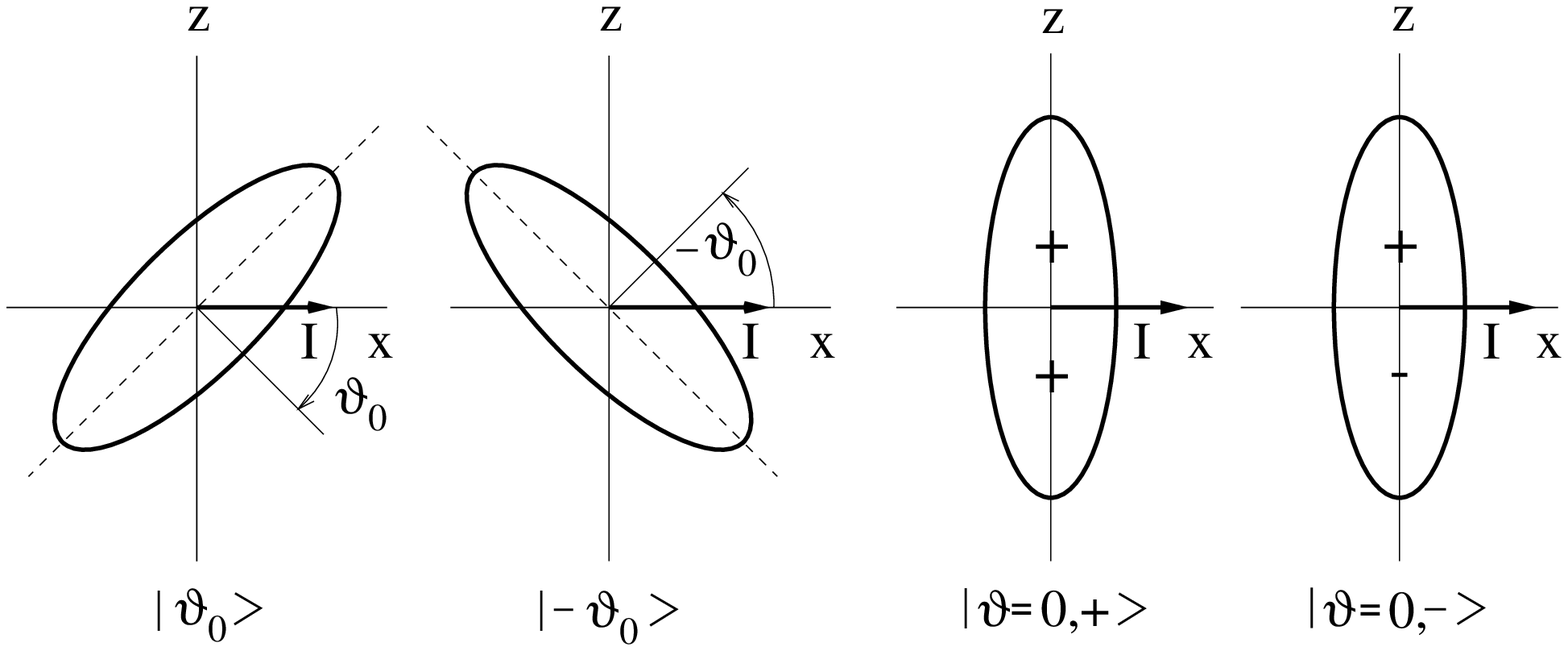,width=14cm,angle=0}}
\caption{\small\it Sketch of the density distrubution of the
four selected \qp\ states after rotation to the lab frame.
The different signature
of the PAC states ( $|\th=0,\pm>$ ) is symbolized as (+,+) and (+,-).
Note, the TAC \qp\ states ( $|\pm\th_\circ>$ ) belong evidently
to two differently oriented mean potentials.
}
\end{figure}

As a result of the cranking term $\om I_x$, the common spin orientation
of all included qp states  is made parallel to the x axis.
 The rotation of a tilted intrinsic state
$|\th>$ reads explicitly as  
\beq
|\th_{lab}> =e^{{-i\th I}_y}\,|\th_\circ> 
\eeq    
where $I_y$ is the y-component of the angular-momentum  operator.
The corresponding density distribution considered in the lab frame of 
reference is sketched  in fig.\ 2. 
Note that the 
tilted qp state changes its density in space 
for the signature operation 
i.e.\ while rotating 180 degrees about the common x-axis. 
This common spin orientation along the x axis  
is taken lateron also as the quantization axis 
for the electromagnetic transition operators. 

For these explorative studies we did not try to precisely match  
the interaction strengths, nor run the qp states to selfconsistent
deformations. Since a quite flexible computer code is available,
several improvements can be made e.g.\  the inclusion of particle
number projection and angular momentum 
projection in x direction by rotating
the qp states in gauge space and about the space-fixed x axis. 
It is intended to build into the code  also a hexadecupole interaction,
in oreder to account for a possible $\epsilon_4$ deformation
of the qp states.

{\bf 4. Results}

The experimental and calculated Routhian energies 
$E'(\om)$ are shown in fig.\ 3 for the frequency interval 
$\om$ = 0.1 to 0.3 MeV.  The  experimental frequency and Routhian are 
obtained from the data with the prescriptions
according to TAC \cite{tac,Meng}.    
The theoretical curves correspond to  TAC, PAC and DISQ approaches,
 using the same parameters as given previously.
To make the comparison with the experimental Routhian better visible, 
we apply 
an arbitrary common shift of the theoretical Routhians  in fig.\ 3.  
\begin{figure}[h]
\begin{minipage}[b]{7cm}
\hspace*{1cm}
\mbox{\psfig{file=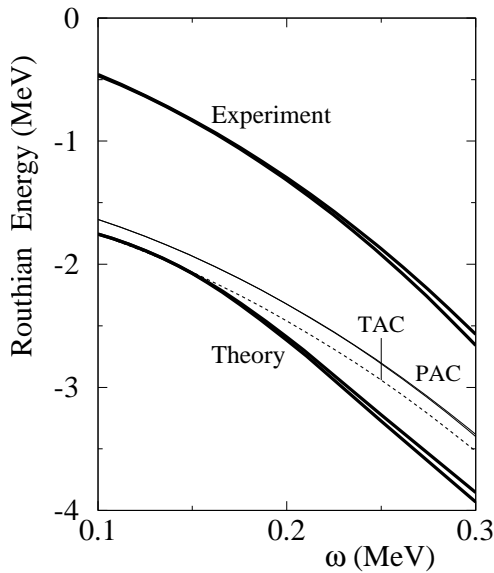,width=7cm,angle=0}}
\end{minipage}
\hspace*{1.5cm}
\begin{minipage}[b]{7cm}
\caption{ \small\it Calculated and experimental
 Routhians for the K=7/2 band in $^{175}$Hf
for frequencies up to 0.3 MeV. In order to be separated from
the experimental curve, all the calculated curves have been
shifted down as a whole. Among them, thin solid lines are from PAC,
dashed ones from TAC and the thick solids from DISQ.
}

\vspace{2cm}

\end{minipage}
\end{figure}
The mixing of the four cranking states described above can nicely reproduce
the experimental splitting of the Routhian into the two signature branches.   
The analogous  signature splitting of the PAC Routhian (i.e.\ the energy 
difference between the two
 quasiparticle states $|\th=0,\pm\rangle $ in fig.\ 2) 
becomes too small by about a factor three.\footnote{Including a positive 
hexadecapole deformation $\epsilon _{4}$ 
in the PAC calculation (not considered here) may increase the
amount of signature splitting, but it is beyond the scope of this paper
to adjust the parameters for the best fit.}
The TAC Routhian belongs to the 
energetically favored quasiparticle states (fig.\ 1) compared to the 
PAC states but without a signature effect. 
Finally, the mixing of the PAC and TAC states leads to an increasing 
signature splitting 
that reaches at $\hbar\om = 0.3$ MeV a few hundred
 keV and a size comparable to the observed splitting.\\  

In fig.\ 4  calculated B(M1) and B(E2) values are displayed as a 
function of the frequency 
$\om$, as found with TAC and PAC quasiparticle states 
and the eigenstates of the DISQ approach. The 
B(E2) values are calculated straightforwardly with the electric quadrupole
operator.  
Concerning the M1 strength, it is known \cite{Hamamoto,Hamamoto1,Meng},
that the calculation of the $\Delta I=1$ magnetic dipole transition 
requires a correction for the nuclear recoil effect similar to the 
center of mass correction in the electric dipole transition. Therefore, the 
effective transversal ($\Delta I=1$) magnetic dipole operator reads as 
\cite{Hamamoto}
\beq
(\mu_{\Delta I=\pm 1})_{eff} = \mu_{\pm 1} - g_{r} I_{\pm 1}
 \eeq 
 where  $g_{r}$ is the so-called gyromagnetic factor of the recoiling
core and $\mu_{\pm 1}$ denote the usual magnetic dipole operator
in spherical representation, i.e.\ for the (lab) x-quantization axis 
one has  
 $\mu_{\pm 1} = \mp 1/\sqrt{2}(\mu_y \pm i\mu_z)$ and accordingly for the 
transversal spin components $I_{\pm 1}$. 

In order to apply a g$_r$-value which is consistent with our 
cranking quasiparticle states, we take as in \cite{Mosel,Ansari}: 
\beq
g_r = \frac{\langle \mu_x \rangle }{\langle I_x \rangle}.      
\eeq 
The $M1$-transition amplitude for pure TAC states $|\th\rangle$ reduces 
to a simple expectation value $\langle\th|\mu_{\pm 1}|\th\rangle $. Then,
 the recoil contribution vanishes automatically 
since the stablility condition (eq.\ 4) for the tilt angle $\th$ 
implies a zero perpendicular spin component  
$\langle I_{\pm} \rangle$ \cite{tac}. This is not any more valid for
the PAC and mixed states where we calculate  the transition  amplitude
between signature partners  as  
$\langle\alpha =-1/2 |\mu_{+1}|\alpha =+1/2\rangle$. The above recoil term
is important for obtaining the right average size of the B(M1) values. However,
it only weakly influences the signature effects and it is not
the origin of  the resulting bifurcations in fig.\ 4. 

For both the B(M1) and B(E2) values,   one realizes in fig.\ 4  
the deficiencies
of the pure mean-field cranking states irrespective of whether one relies on 
TAC or PAC. 
The TAC
approach gives the correct geometrical dependence of the M1 and E2 
transition rates. Qualitatively, the tilt angle $\th$ is large for the lowest
frequency and, correspondingly,  the rotational axis (spin orientation) is
relatively close to the symmetry axis. Therefore, for low $\om$ 
both the B(M1) and the B(E2) values are expected to be small 
since the effective deformation and the perpendicular magnetic moment
seen along this rotational axis are small. 
For larger frequencies $\om$,
the tilt angle $\th$ approaches zero and the rotational axis becomes
more and more perpendicular
to the symmetry axis, which leads to increasing effective deformation and 
perpendicular magnetic moment, i.e.\ to ascending 
B(M1) and B(E2) strength. 
Hence, this general frequency dependence is reflected by the TAC approach.
This is not the case for the PAC states where in particular the B(E2)
becomes constant. However, the PAC treatment can describe correctly 
the development of the signature effect in the B(M1) strength,
which is outside the range of TAC.

The mixing of TAC and PAC states within DISQ indeed can reproduce both the 
geometrical dependence and the signature bifurcation. 
Thus, this model achieves
the goal for which it was designed. In particular, for the B(M1) strength
one realizes the continuous transition from the TAC to the PAC regime when
increasing the rotational frequency.

\begin{figure}[h]
\begin{minipage}[b]{7cm}
\hspace*{0.8cm}
\mbox{\psfig{file=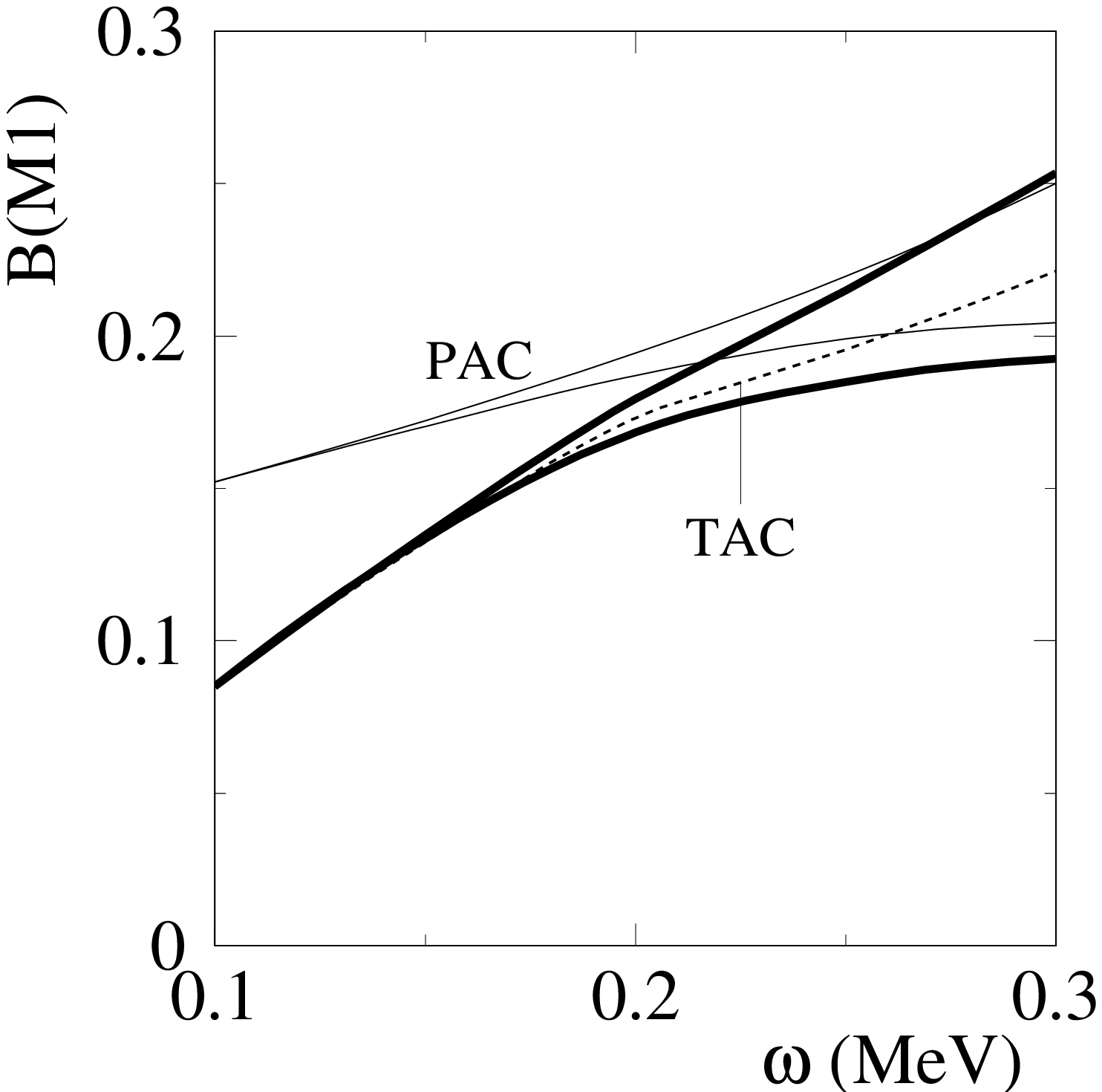,width=7cm,angle=0}}
\end{minipage}
\hspace*{1.3cm}
\begin{minipage}[b]{7cm}
\mbox{\psfig{file=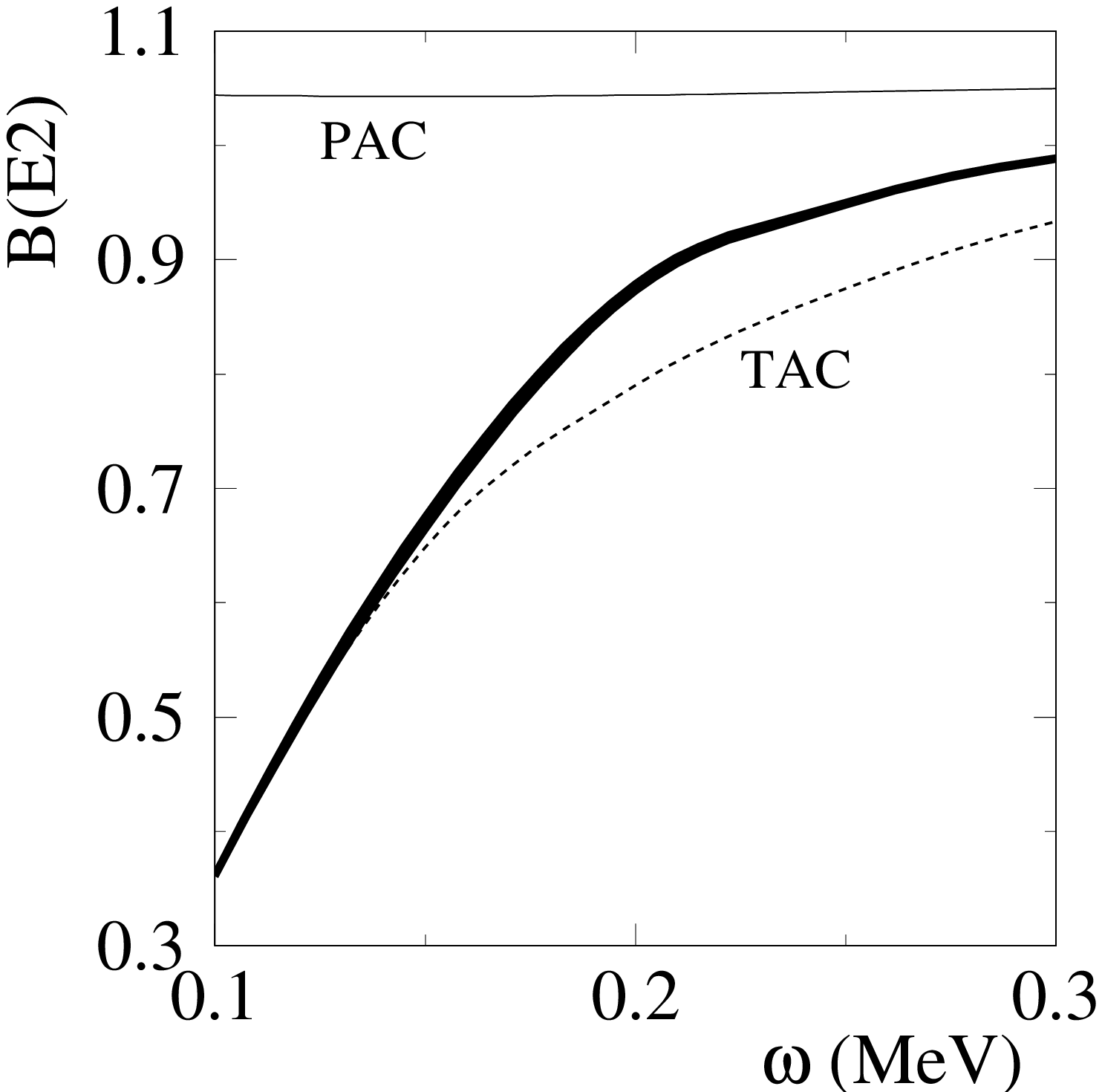,width=7cm,angle=0}}
\end{minipage}
\hspace*{-0.cm}
\caption{ \small\it Calculated B(M1) in ($\mu_N$)\,$^2$ (left) and B(E2)
values in (eb)$^2$ (right).
Dashed curves: TAC, thin solid curves: PAC, thick solid curves: DISQ.
The B(M1) values show the developing signature bifurcation
which is consistent with the energy signature splitting
seen in fig\ 3.
There is no signature effect in the B(E2) values since
the quadrupole deformation of two signature partners is about the same.
}
\end{figure}

{\bf 5. Summary }

The proposed DISQ approach, i.e.\ basically the superposition of appropriate   
mean-field quasiparticle states by diagonalization, is found to heal 
the apparent deficiences of the mean-field cranking model
related to the signature symmetry. 
This microscopic approach
accomplishes the description of the gradually growing admixture 
of the intrinsic signature oscillation to the rotational
motion seen experimentally as an expanding
bifurcation of the band energies and in parallel in the B(M1)  
transition strength. This splitting  is a signal for the smooth transition 
from the TAC to the PAC regime, which is definitely out
of the reach of the mean-field approximation.

The actual calculations performed for $^{175}$Hf included the total 
Routhian energy as well as 
the M1 and E2 transition strength
as a function of frequency in the
interval $\hbar\om = 0.1 - 0.3 $ MeV.
The expected
trends for the increasing development of the signature
splitting and simultaneous
staggering of the B(M1) transitions between the two
signatures partners can be nicely described within our approach. 

The signature restoration in an odd-A nucleus considered 
in this paper is a relatively simple case of restoring the 
signature symmetry. 
However, the 
DISQ method is intended to be applied also 
to more complex situations, e.g.\ to odd-odd nuclei
and triaxial nuclei where the bifurcation patterns are more 
complicated (c.f.\ \cite{signat}). For triaxial shapes 
additional stability points exist in the corresponding potential 
energy surface opening the possibility of novel tunneling modes
between them. 
The study of those situations will be a new 
interesting subject. The DISQ approach is, of course, a general method which
might be appropriate for other types of large amplitude motion,  e.g.\
 $K$-ismeric decay and band-crossing phenomena.  
 
In a more general context, the symmetry restoration by the DISQ approach 
can be considered as a successful example for treating the many-body 
quantum motion in a double humped potential. The splitting process
resembles the picture of a phase transition in a finite system where the rotational 
frequency plays the role of an order parameter. At low $\om$ values with practically 
vanishing splitting, the nucleus can stay in a symmetry-broken phase 
having accidentally one of the stable orientations $\pm\th_\circ$. 
For increasing $\om$ values the system more and more bifurcates in two phases,
realizing the required D2 symmetry. There is obviously a continuous transition region
between the signature-broken TAC and the signature-conserving PAC 
regime.

{\bf Acknowledgements.} This work is supported by the U.~S. Department
                            of Energy through contract no.\
                            DE-FG05-96ER40983 
and by German Federal Ministry of Education, Science, Research and Technology. We like 
also to thank  S.\ Frauendorf for important discussions.

\end{document}